\documentclass[a4paper, times, 10pt, twocolumn, twoside,top=3.5cm,bottom=3.5cm,left=2.5cm,right=2cm]{article}

\usepackage{ISARC}
\usepackage{graphicx,subfigure}
\usepackage{color}
\usepackage{CJK}
\usepackage{framed}
\usepackage{ulem}

\usepackage{caption}
\usepackage{amsmath}
\usepackage{commath}

\begin{document}

\renewcommand{\ISARCConference}{}

\definecolor{shadecolor}{rgb}{1,1,0}
\linespread{0.5}
\title{Adaptive Second-order Sliding Mode Control of UAVs \\for Civil Applications}

\author{V.T. Hoang, A.M. Singh, M.D. Phung and Q.P. Ha}
\affiliation{Faculty of Engineering and Information Technology\\
University of Technology Sydney, Australia}
\email{\{VanTruong.Hoang,  AnsuMan.Singh,
ManhDuong.Phung, Quang.Ha\}@uts.edu.au}

\maketitle 
\thispagestyle{fancy} \pagestyle{fancy}

\begin {abstract}
Quadcopters, as unmanned aerial vehicles (UAVs), have great potential in civil applications such as surveying, building monitoring, and infrastructure condition assessment. Quadcopters, however, are relatively sensitive to noises and disturbances so that their performance may be quickly downgraded in the case of inadequate control, system uncertainties and/or external disturbances. In this study, we deal with the quadrotor low-level control by proposing a robust scheme named the adaptive second-order quasi-continuous sliding mode control (adaptive 2-QCSM). The ultimate objective is  for robust attitude control of the UAV in monitoring and inspection of built infrastructure. First, the mathematical model of the quadcopter is derived considering nonlinearity, strong coupling, uncertain dynamics and external disturbances. The control design includes the selection of the sliding manifold and the development of quasi-continuous second-order sliding mode controller with an adaptive gain. Stability of the overall control system is analysed by using a global Lyapunov function for convergence of both the sliding dynamics and adaptation scheme. Extensive simulations have been carried out for evaluation. Results show that the proposed controller can achieve robustness against disturbances or parameter variations and has better tracking performance in comparison with experimental responses of a UAV in a real-time monitoring task.
\end{abstract}


\section{Introduction}
Quadcopters have found many applications in civil engineering automation due to its flexibility in operational space and ability in vertical take off and landing. These include the use of UAVs in automatic 3D reconstruction for building condition assessment \cite{Chen:2015}, securing superstructures of high-rise buildings \cite{Choi:2015}, or monitoring and inspection of civil infrastructure \cite{Ham:2016, Phung:2016}. In those applications, it is critical to maintain robustness and resilience of the control system to cope with the highly non-linear dynamics of quadcopters and system uncertainties, sensor noise and coupling effects between the rotational and translational motions, or disturbances from aerodynamics and other external factors.

A number of control approaches have been developed for the quadcopter in the literature, for example PD, PID control \cite{Zuo:2010}, $H_\infty$ control \cite{Raffo:2010}, optimal control \cite{Ritz_2011}, or potential field \cite{La_2016}. Among them, the sliding mode control (SMC) is widely used as it can produce a robust closed-loop control system under the influence of modelling errors and external disturbances \cite{Xu:2006,Besnard:2007,Derafa:2012}. In SMC, chattering may occur in the steady state and act as an oscillator that excites unmodeled frequencies of the system dynamics \cite{Manceur:2012}. 
To reduce the chattering effect, high-order sliding modes (HOSM) have been introduced \cite{Levant:1993, Polyakov:2009, Rubio:2014, Utkin:2016}. 

In the HOSM control, the quasi-continuous (QC) SMC \cite{Ding:2015} introduces the capability of maintaining the properties of the first order SMC while creating smooth responses. Its performance however depends on the knowledge of disturbance boundaries which are not always available. In practice, the quadcopter may be subject to various disturbances and uncertainties such as wind gusts and modelling errors that may downgrade the control performance. To address this concern, the second-order sliding mode (SOSM) controller with an adaptive gain has been applied to drive the sliding variable and its derivative to zero in the presence of bounded disturbances \cite{Shtessel:2012}.

In this work, we propose an adaptive quasi-continuous second-order sliding mode (AQCSM) scheme to control the attitude of quadcopters subject to nonlinear dynamics, strong coupling, high uncertainties and disturbances with unknown boundaries. The mathematical model of the quadcopter is first derived by considering various dynamic parameters. Here, the quasi-continuous SMC retains the advantage of robustness while attenuating the control chattering and facilitating the implementation. Its performance is verified by simulation with comparison to real-time datasets.

The paper is organised as follows. The dynamic model of the quadcopter is presented in Section 2. Section 3 describes the development of the AQCSMC. Simulation results are presented in Section 4 with comparison to PID experimental responses. The paper ends with a conclusion and discussion for future work.

\section{System modelling}
\subsection{Kinematics}
\begin{figure}[!htbp]
\centering
\includegraphics[width=8.2cm]{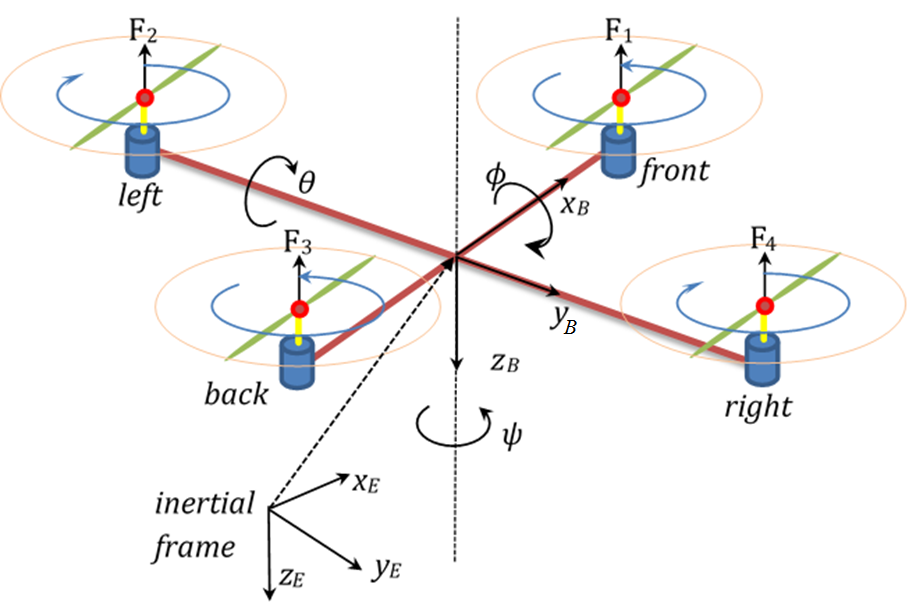}
\caption{A schematic diagram of quadcopter}
\label{fig_1}
\end{figure}

Two coordinate systems are used to model the kinematics and dynamics of quadrotors, as shown in Fig. \ref{fig_1}. The inertial frame $(x_E, y_E, z_E)$ is defined by the ground with gravity pointing downward in $z_E$ direction. The body frame $(x_B, y_B, z_B)$ is specified by the orientation of the quadcopter with the rotor axes pointing in the positive $z_B$ direction and the arms pointing in $x_B$ and $y_B$ directions.

The orientation of quadcopters is described by the roll, pitch, and yaw angles corresponding to its rotations around the $x_B$, $y_B$ and $z_B$ axes. Denoting those angles as  $\Theta = ({\phi , \theta , \psi})^T$, their rates are then given by $\dot \Theta =  ({\dot \phi , \dot \theta , \dot \psi})^T$. The rates relate with angular velocities, $\omega = [p,q,r]^T$, by the following transformation:
\begin{equation}\label{Eq1}
\omega = H\dot\Theta,
\end{equation}
where $H$ is given by:
\begin{equation}\label{Eq1b}
H= \left [\begin{array}{ccc} 
    1 &0 & -s_\theta \\
    0 & c_\phi & c_\theta s_\phi \\
    0 & -s_\phi & c_\theta c_\phi 
\end{array} \right ],
\end{equation}

\noindent in which $s_x = sin(x)$ and $c_x = cos(x)$. As the result, the rotational matrix of the quadcopter is described by:
\begin{equation}\label{Eq2}
R = \left [\begin{array}{ccc} 
c_\psi c_\theta & c_\psi s_\theta s_\phi-s_\psi c_\phi & c_\psi s_\theta c_\phi+s_\psi s_\phi\\
s_\psi c_\theta & s_\psi s_\theta s_\phi+c_\psi c_\phi & s_\psi s_\theta c_\phi-c_\psi s_\phi\\
-s_\theta & c_\theta s_\phi & c_\theta c_\phi
\end{array} \right].
\end{equation}
\subsection{Quadcopter Dynamics}
Since the focus is on the attitude control so only torque components that cause changes in the orientation are considered. They include torques caused by thrust forces $\tau$, body gyroscopic effects $\tau_b$, propeller gyroscopic effects $\tau_p$, and aerodynamic friction $\tau_a$. The torque $\tau$ consists of three components corresponding the roll, pitch and yaw rotations, $\tau = [\tau_\phi \enskip \tau_\theta \enskip  \tau_\psi]^T$. They are given by:
\begin{equation}
\tau_\phi = l(F_2 - F_4),
\label{eq:torque_roll}
\end{equation}
\begin{equation}
\tau_\theta = l(-F_1 + F_3),
\label{eq:torque_pitch}
\end{equation}
\begin{equation}
\tau_\psi = b(-F_1+F_2 -  F_3+F_4),
\label{eq:torque_yaw}
\end{equation}
where $l$ is the distance from the motor to the UAV centre of mass and $b$ is the drag factor. The body gyroscopic torque $\tau_b$ is given by:
\begin{equation}
\tau_b = -S(\omega) I\omega,
\end{equation}
where $S(\omega)$ is a skew-symmetric matrix,
\begin{align}\label{Eq5a}
S(\omega) = \left[ \begin{array}{ccc}
0 & -r & q\\ r & 0 & -p\\ -q & p & 0
\end{array}\right].
\end{align}

The propeller gyroscopic torque $\tau_p$ is determined as:
\begin{equation*}
\tau_p = \left[ \begin{array}{c}
I_r \Omega_r q\\
-I_r \Omega_r p\\
0
\end{array} \right],
\end{equation*}
where $I_r$ is the inertial moment of rotor, $\Omega_r = -\Omega_1 + \Omega_2 - \Omega_3 + \Omega_4$ is the residual angular velocity of rotor in which $\Omega_k$ denotes the angular velocity of the propeller $k$ ($k$=1,2,3,4). Finally, the aerodynamic friction torque $\tau_a$ is given by:
\begin{equation}
\tau_a = k_a \omega^2,
\end{equation}
where $k_a$ depends on aerodynamic friction factors, $k_a = [k_{ax}, k_{ay}, k_{az}]^T$. Given those torque components, the attitude dynamic model of the quadcopter is described as:
\begin{equation}\label{Eq5}
I\ddot{\Theta} = \tau_b  + \tau + \tau_p - \tau_a,
\end{equation}
where $I = \text{diag}[I_{xx},  I_{yy}, I_{zz}]$ is the inertia matrix when the quadrotor is assumed to be symmetrical. 

In our system, the gyroscopic and aerodynamic torques are considered as external disturbances. Thus, the control inputs mainly depend on torque $\tau$ and from (\ref{eq:torque_roll}), (\ref{eq:torque_pitch}) and (\ref{eq:torque_yaw}), they can be represented as:
\begin{align}\label{Eq6}
\left[\begin{array}{c} u_\phi\\ u_\theta\\ u_\psi \\ u_z\end{array}\right] = \left[\begin{array}{c} \tau_\phi\\ \tau_\theta\\  \tau_\psi \\ F\end{array}\right]  = \left[\begin{array}{cccc}
0 & l & 0 & -l \\	-l & 0 & l & 0\\ -c & c & -c & c \\ 1 & 1 & 1 & 1
\end{array}\right] \left[\begin{array}{c} F_1 \\ F_2\\ F_3\\ F_4\end{array}\right],
\end{align}
where $F=F_1+F_2+F_3+F_4$ is the UAV lift, $u_z$ represents the total thrust acting on the four propellers and $u_\phi$, $u_\theta$ and  $u_\psi$ respectively represent the roll, pitch and yaw torques, $c$ is a force-to-torque scaling coefficient. As only the attitude of quadcopter will be controlled, $u_z$ is assumed to balance with  the gravity. Consequently, the dynamics of quadcopters can be represented in the following form for attitude control:
\begin{equation}\label{Eq7}
\dot{\omega} = I^{-1}\left( -S(\omega) I\omega + U + d\right), 
\end{equation}
where $U = [u_\phi, u_\theta, u_\psi]^T$ is the input vector and $d = [d_\phi, d_\theta, d_\psi]^T$ is the disturbance vector. In our system, the following assumptions are made:
\begin{itemize}
\item[A.1] The quadcopter structure is rigid and symmetric. The propellers are rigid.
\item[A.2] The signals $\Theta$ and $\dot{\Theta}$ can be measured by on-board sensors.
\item[A.3] The reference trajectories and their first and second time derivatives are bounded.
\item[A.4] The velocity and the acceleration   of the quadcopter are bounded.
\item[A.5] The orientation angles are limited to $\phi \in \left[-\dfrac{\pi}{2}, \dfrac{\pi}{2}\right]$, $\theta \in \left[-\dfrac{\pi}{2}, \dfrac{\pi}{2}\right]$ and $\psi \in \left[-\pi, \pi\right]$.
\item[A.6] The rotational speeds of rotors are bounded.
\end{itemize}
\section{Control Design}
The control signals $u_\phi, u_\theta$ and $u_\psi$ in (\ref{Eq7}) are used to control the three angles $\{\phi, \theta, \psi\}$ to reach the reference value $\Theta_{d} = \{\phi_d, \theta_d, \psi_d\}^T$.

\subsection{Sliding Manifold} \label{SMd}
The sliding function determining the system's equivalent dynamics is presented as:
\begin{equation}\label{Eq10}
\mathbf{\sigma} = \dot{\mathbf{e}} + \Lambda \mathbf{e},
\end{equation}
where $\Lambda = \text{diag}(\lambda_\phi, \lambda_\theta, \lambda_\psi)$ is a positive definite matrix to be designed, and $\mathbf{e} = \Theta_{d} - \Theta$ is the control error. Taking the derivative of $\sigma$, we have:
\begin{align}\label{Eq14}
\dot{\sigma} &=  \ddot{\Theta} - \ddot{\Theta}_{d} + \Lambda\dot{\mathbf{e}}.
\end{align}
For small angular rotations of the quadcopter, we can approximate $\omega$ to $\dot{\Theta}$ \cite{Zheng:2014}. Substituting $\ddot{\Theta}$ (\ref{Eq7}) to (\ref{Eq14}) yields: 
\begin{equation}\label{Eq15}
\dot{\sigma} = -\ddot{\Theta}_{d} + \Lambda\dot{\mathbf{e}} + I^{-1}[-S(\omega) I\omega + U + d].
\end{equation}

\subsection{QCSM control design and problem formulation} \label{SMd}
The second-order sliding mode control proposed in \cite{Levant:2003,Levant:2007} is used in this paper, for which a conventional QCSM is defined as follows:
\begin{equation}\label{Eq19}
U = -\alpha \dfrac{\dot{\sigma} + \abs{\sigma}^{1/2}\text{sign}(\sigma)}{\abs{\dot{\sigma}} + \abs{\sigma}^{1/2}},
\end{equation}
where $\alpha$ is the control gain to be adjusted. The control is continuous everywhere apart from the origin where $\sigma = \dot{\sigma} = 0$.


Since $I$ is symmetric and positive definite, the following Lyapunov function is chosen to avoid the inversion of the inertia matrix:
\begin{equation}\label{Eq19a0}
V_0 = \dfrac{1}{2}\sigma^T I \sigma.
\end{equation} 

Taking the time derivative of $V$ gives 
\begin{equation}\label{Eq19a1}
\dot{V}_0 = \dfrac{1}{2}\left( \dot{\sigma}^T I \sigma + \sigma^T I \dot{\sigma}\right) +\dfrac{1}{2} \sigma^T \dot{I} \sigma = \sigma^T\left(I \dot{\sigma} + \dfrac{1}{2}  \dot{I} \sigma\right) .
\end{equation}
By substituting $\dot{\sigma}$  from (\ref{Eq19a1}) to (\ref{Eq15}), one has
\begin{equation}\label{Eq19a}
\dot{V}_0 = \sigma^T\left( -I\ddot{\Theta}_{d} + I\Lambda \dot{e}-S(\omega) I\omega + U + d + \dfrac{1}{2} \dot{I}\sigma \right).
\end{equation}

Let $I = I_0 + \Delta I$, where $I_0$ and $\Delta I$ represent the nominal and uncertain parts of the inertia matrix. According to A1, we have $\dot{I} = 0$, equation (\ref{Eq19a}) becomes
\begin{align}\nonumber
\dot{V}_0 = \sigma^T \{ & -S(\omega) \Delta I\omega- \Delta I\ddot{\Theta}_{d} +\Delta I\Lambda \dot{e} + d+ \dfrac{1}{2} \dot{I}\sigma\\
& + U - S(\omega) I_0\omega - I_0\ddot{\Theta}_{d}+ I_0\Lambda \dot{e} \\ \label{Eq19b}
= \sigma^T \{ & \Delta P + U + P \}, 
\end{align}
where
\begin{align}\label{Eq19b1}
\Delta &P = -S(\omega) \Delta I\omega- \Delta I\ddot{\Theta}_{d} + \Delta I\Lambda \dot{e} + d,\\\label{Eq19b2}
&P = -S(\omega) I_0\omega - I_0\ddot{\Theta}_{d}+ I_0\Lambda \dot{e}.
\end{align}
Let $\Xi=[\Xi_1, \Xi_2,\Xi_3]^T$ denote the sum of $\Delta P$ and $P$. Since the disturbance $d$ and uncertain parameter $\Delta I$ are bounded, from (\ref{Eq19b1}) and (\ref{Eq19b2}) it can be seen that  $\Xi$ is also bounded, i.e., $|\Xi_i| \le \Xi_{M,i}, ~i=1,2,3$. 
Consider system (\ref{Eq7}) with the sliding variable $\sigma(\omega,t)$ as in (13). From assumptions A1-A6, the sliding motion on the manifold is achieved by the controller (\ref{Eq19}) if we can select the gain $\alpha_i$ such that \cite{pukdeboon:2010}:
\begin{equation}
\alpha_i \geq \Xi_{M,i}.
\end{equation}
However, the bound $\Xi_{M,i}$ is not easy to evaluate in practice and besides, there is a trade-off with chattering if a high value of $\alpha_i$ is chosen. The problem is now to drive the sliding variable $\sigma$ and its derivative $\dot{\sigma}$ to zero in finite time by means of quasi-continuous SMC without  overestimation of the control gain. 

\subsection{Adaptive QCSM Design}
The proposed gain-adaptation law is supposed to minimise the chattering phenomenon while driving $\sigma$ and $\dot{\sigma}$ to zero 
even in the presence of disturbances. For initial conditions $\omega_i(0), \sigma_i(0)$, and $\alpha_i(0) > 0$, the reaching and sliding on the manifold is globally achieved in finite time by the controller (\ref{Eq19}) with the following adaptive gain \cite{Plestan:2010}: 
\begin{align}\label{Eq24a12}
\dot{\alpha}_i &= \begin{cases}
\bar{\omega}_i \abs{\sigma_i(\omega,t)} \text{sign}(|\sigma_i(\omega,t)|-\epsilon_i) &\text{if} \enskip \alpha_i > \alpha_{m,i}\\
\eta_i &\text{if} \enskip\alpha_i \leq \alpha_{m,i},
\end{cases}
\end{align}
where  $\bar{\omega}_i >0$, $\epsilon_i$, $\eta_i$ are small positive constants, and $\alpha_{m,i}$ is a threshold of the adaptation. \\
%

To analyse the stability of the proposed controller, let us first define a global Lyapunov function candidate for $\sigma$ and $\alpha$ as:
\begin{equation}\label{Eqn20}
V(\sigma, \alpha) = V_0 + \sum\limits_{i=1}^3\dfrac{1}{2\gamma_i}(\alpha_i -\alpha_{M,i})^2,
\end{equation}
where $V_0$ has been defined in Eq. (\ref{Eq19a0}), $\gamma_i$ is some positive constant and $\alpha_{M,i}$ is the maximum possible value of the adaptive gain $\alpha_i$. The derivative of the Lyapunov function (\ref{Eqn20}) is obtained as
\begin{equation}\label{Eqn21}
\dot{V}(\sigma, \alpha) = \dot{V}_0 + \sum\limits_{i=1}^3\dfrac{1}{\gamma_i}(\alpha_i -\alpha_{M,i})\dot{\alpha}_i.
\end{equation}
Taking $\dot{V}_0$ from (2) and $\dot{\alpha}_i$ from (\ref{Eq24a12}), equation (\ref{Eqn21}) under the control law (16) becomes
\begin{align}\nonumber
\dot{V}(\sigma, \alpha) &= \sum\limits_{i=1}^3 \sigma_i \left[\Xi_i   -\alpha_i \left( \dfrac{\dot{\sigma}_i + \abs{\sigma_i}^{1/2}\text{sign}(\sigma_i)}{\abs{\dot{\sigma}_i} + \abs{\sigma_i}^{1/2}}\right) \right]  +\\\label{Eqn21c}
&+ \sum\limits_{i=1}^3\dfrac{1}{\gamma_i}(\alpha_i -\alpha_{M,i}) \bar{\omega}_i \abs{\sigma_i}\text{sign}(|\sigma_i|-\epsilon_i).
\end{align}
When $\sigma_i$ is slowly time-varying, $\dot{\sigma}_i(t)$ is very small and can be negligible, then equation (\ref{Eqn21c}) becomes
\begin{align}\nonumber
\dot{V}(\sigma, \alpha) &= \sum\limits_{i=1}^3 \sigma_i \left[\Xi_i   -\alpha_i \text{sign}(\sigma_i) \right] + \\\label{Eqn21c0}
&+ \sum\limits_{i=1}^3\dfrac{1}{\gamma_i}(\alpha_i -\alpha_{M,i}) \bar{\omega}_i \abs{\sigma_i}\text{sign}(|\sigma_i|-\epsilon_i).
\end{align}
It can be seen that $\dot{V} \le 0$ given (23) and $\alpha_i \le\alpha_{M,i}$ \cite{Plestan:2010}.


\section{Simulation and Validation}
\begin{figure}[!htbp]
\centering
\includegraphics[width=8.2cm]{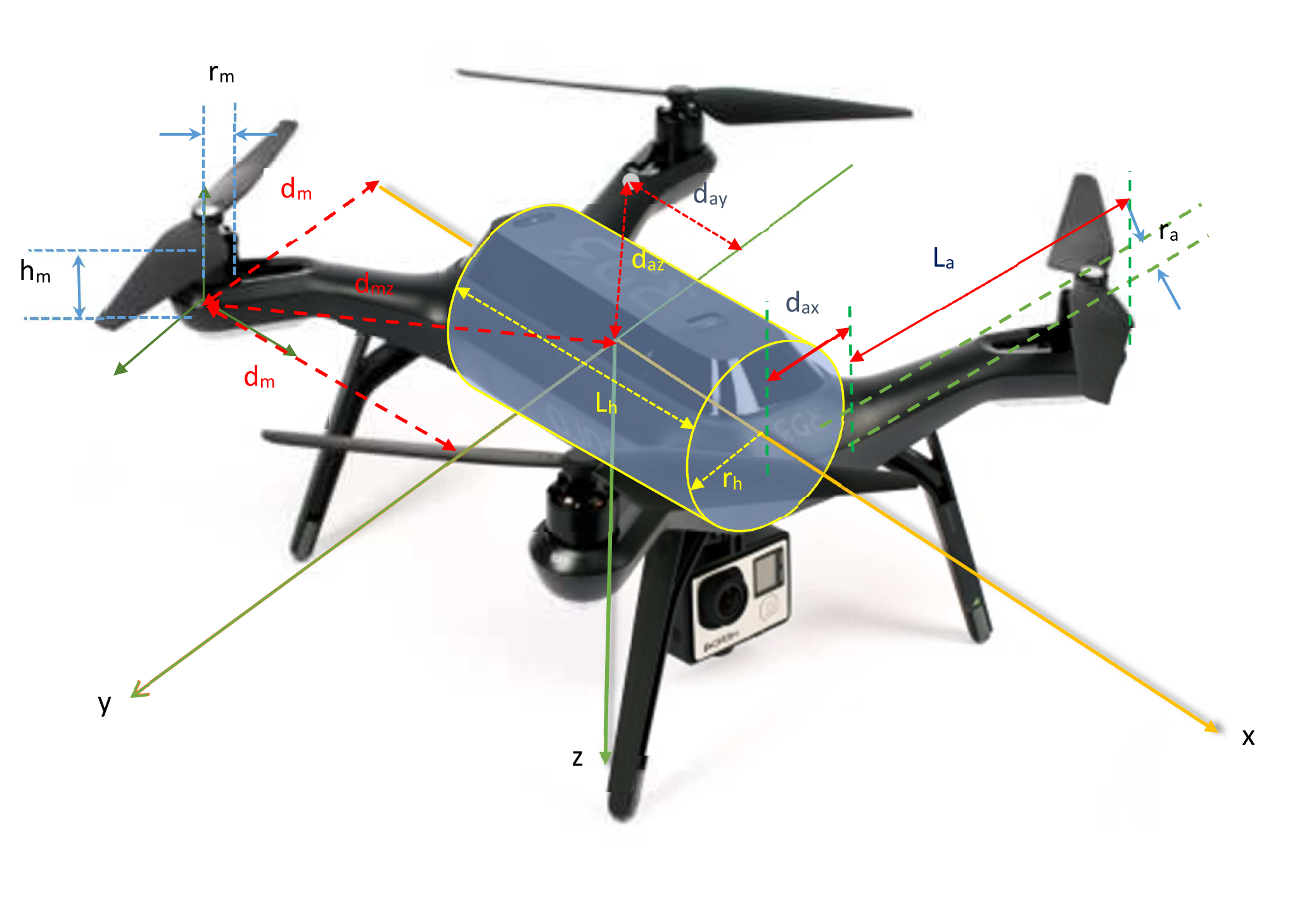}
\caption{The 3DR Solo drone with body coordinate frame.}
\label{FigQuad}
\end{figure}

\begin{figure}[!htbp]
\centering
\includegraphics[width=8.2cm]{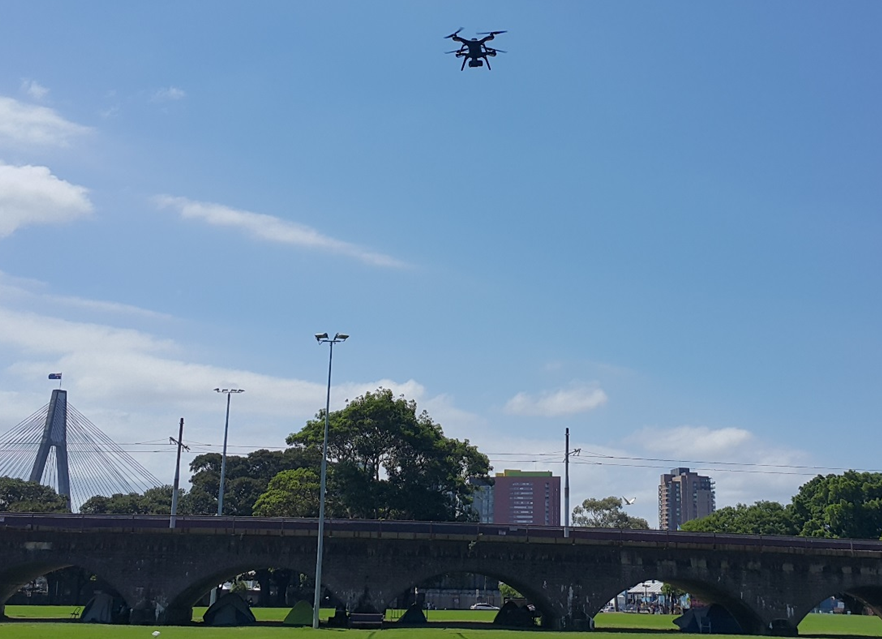}
\caption{Insfrastructure inspection.}
\label{MonQuad}
\end{figure}

Extensive simulation has been carried out to evaluate the performance of the proposed control algorithm. The model of the test quadcopter used is obtained from the 3DR Solo drone shown in Fig. \ref{FigQuad} in which $L_x$, $d_x$, $r_x$ and $h_x$ are measured distances used to compute system parameters, as listed in Table \ref{Table1}. Design parameters used for the controllers are given in Table \ref{Table2}.
The UAV, with technical specifications and accessories described in \cite{Ha_2017}, was deployed to perform the tasks of infrastructure inspection, as shown in Fig. \ref{MonQuad}.

\begin{table}[htbp]
\centering
\caption{Parameters of the quadcopter model }
\begin{tabular}{ccc}
\hline\label{Table1}
Parameter & Value & Unit \\ \hline
$m$ & 1.50 & \textit{kg} \\
$l$ & 0.205 & \textit{m} \\
$g$ & 9.81 & $m/s^2$ \\		
$I_{xx}$ & $8.85\cdot 10^{-3}$& $kg.m^2$\\	
$I_{yy}$ & $15.5 \cdot 10^{-3}$ & $kg.m^2$\\
$I_{zz}$ & $23.09 \cdot 10^{-3}$ & $kg.m^2$\\
\hline
\end{tabular}
\end{table}

\begin{table}[htbp]
\centering
\caption{Control design parameters}
\begin{tabular}{cccc}
\hline\label{Table2} 
Variable & Value & Variable & Value \\ \hline
$\lambda_1$ & 4.68 & $\lambda_2$ & 4.68\\
$\lambda_3$ & 3.84 & $\epsilon_{1,2,3}$ & 0.7\\
$\alpha_0$ & 1.24 & $\bar{\omega}_{1,2,3}$ & 200\\
$\alpha_{m,1}$ & 0.01 & $\alpha_{m,2}$ & 0.02\\
$\alpha_{m,3}$ & 0.03 &$\eta_{1,2,3}$ & 0.01 \\
\hline
\end{tabular}
\end{table}
\subsection{Control performance in nominal conditions}
\begin{figure}[!htbp]
\centering
\includegraphics[width=8.2cm]{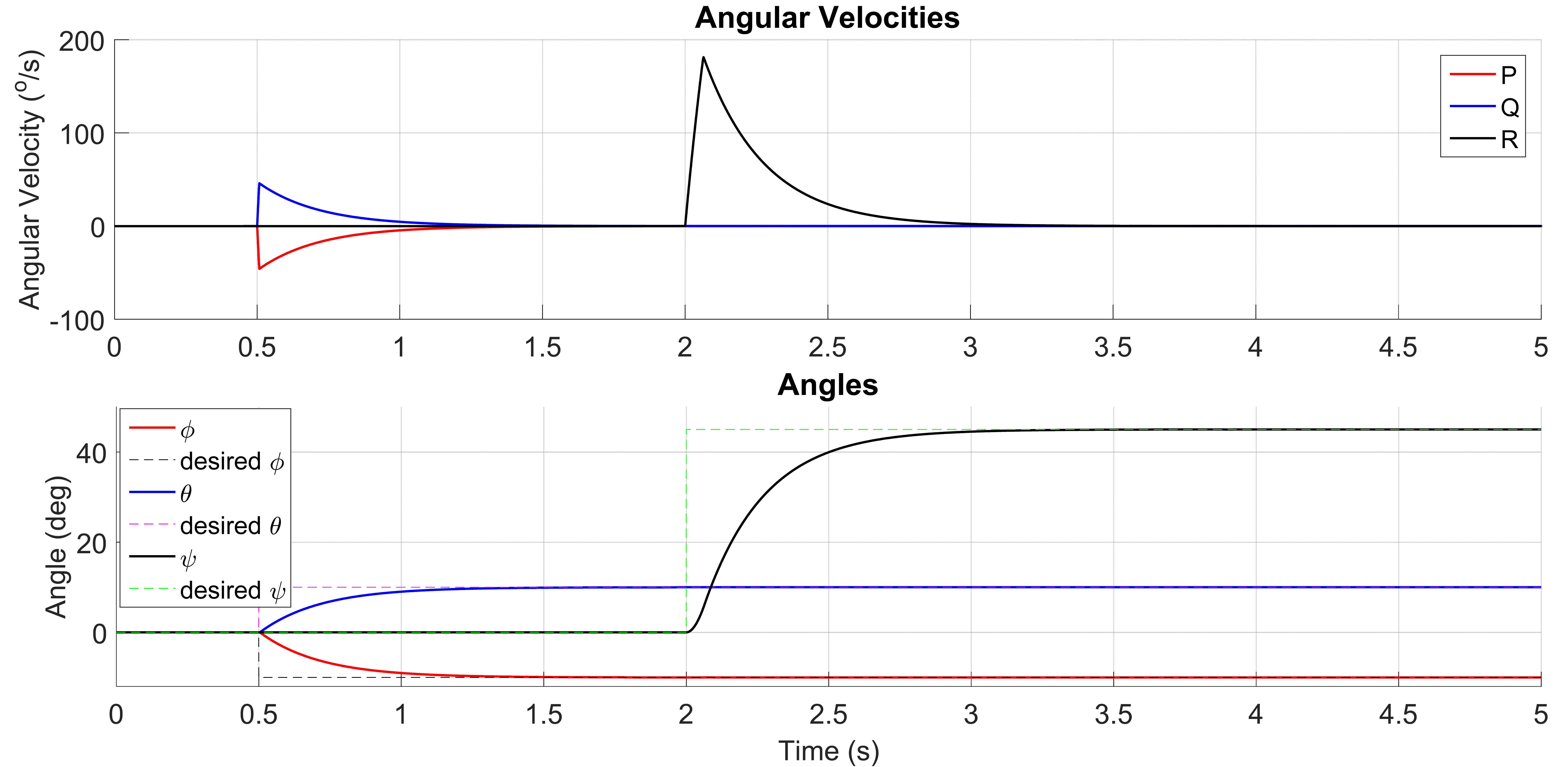}
\caption{Responses of the quadcopter in nominal conditions ($P$, $Q$ and $R$- roll, pitch and yaw angular velocities).}
\label{FigPhi}
\end{figure}

\begin{figure}[!htbp]
\centering
\includegraphics[width=8.2cm]{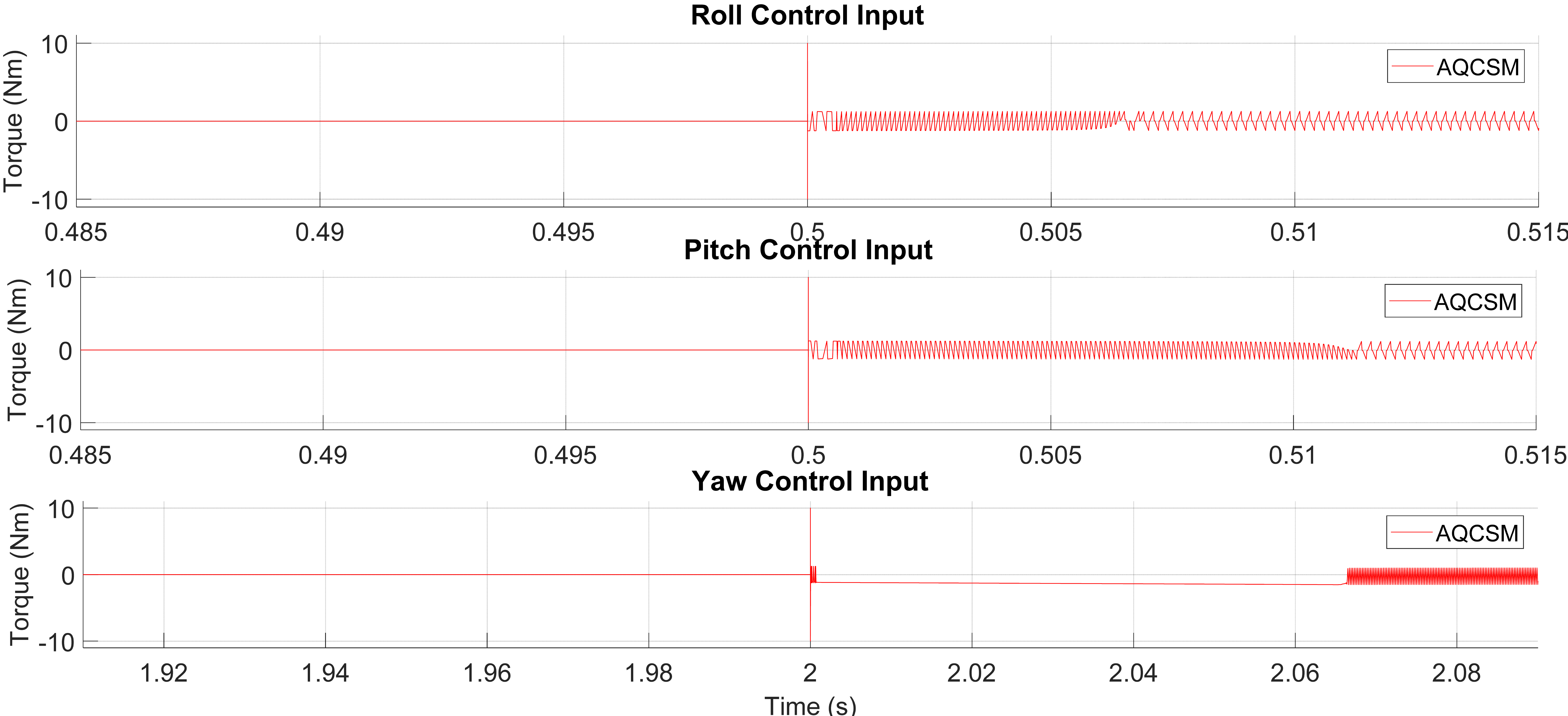}
\caption{Control torques.}
\label{Fig:torques}
\end{figure}

In this simulation, the quadcopter starts from zero initial conditions, i.e. all angles and velocities are zeros. Its roll and pitch angles are then set to $\phi = -10^0$ and $\theta = 10^0$ at time 0.5 s and its yaw angle is then set to $\psi = 45^0$ at time 2 s. The results are shown in Fig. \ref{FigPhi} and Fig. \ref{Fig:torques}, where the time scale in the latter is zoomed in to observe the abrupt change in the control torque and coupling effect. It can be seen that all controllers smoothly drive the angles to the desired values with relatively small overshoot and within two seconds. According to (\ref{Eq6}), there exist strong coupling relations between the control states. As a result, it can been seen that the AQCSM controller can handle this problem to control the attitude to reach the reference values and then track them without being perturbed.

\subsection{Responses to disturbances}
\begin{figure}[!htbp]
\centering
\includegraphics[width=8.2cm]{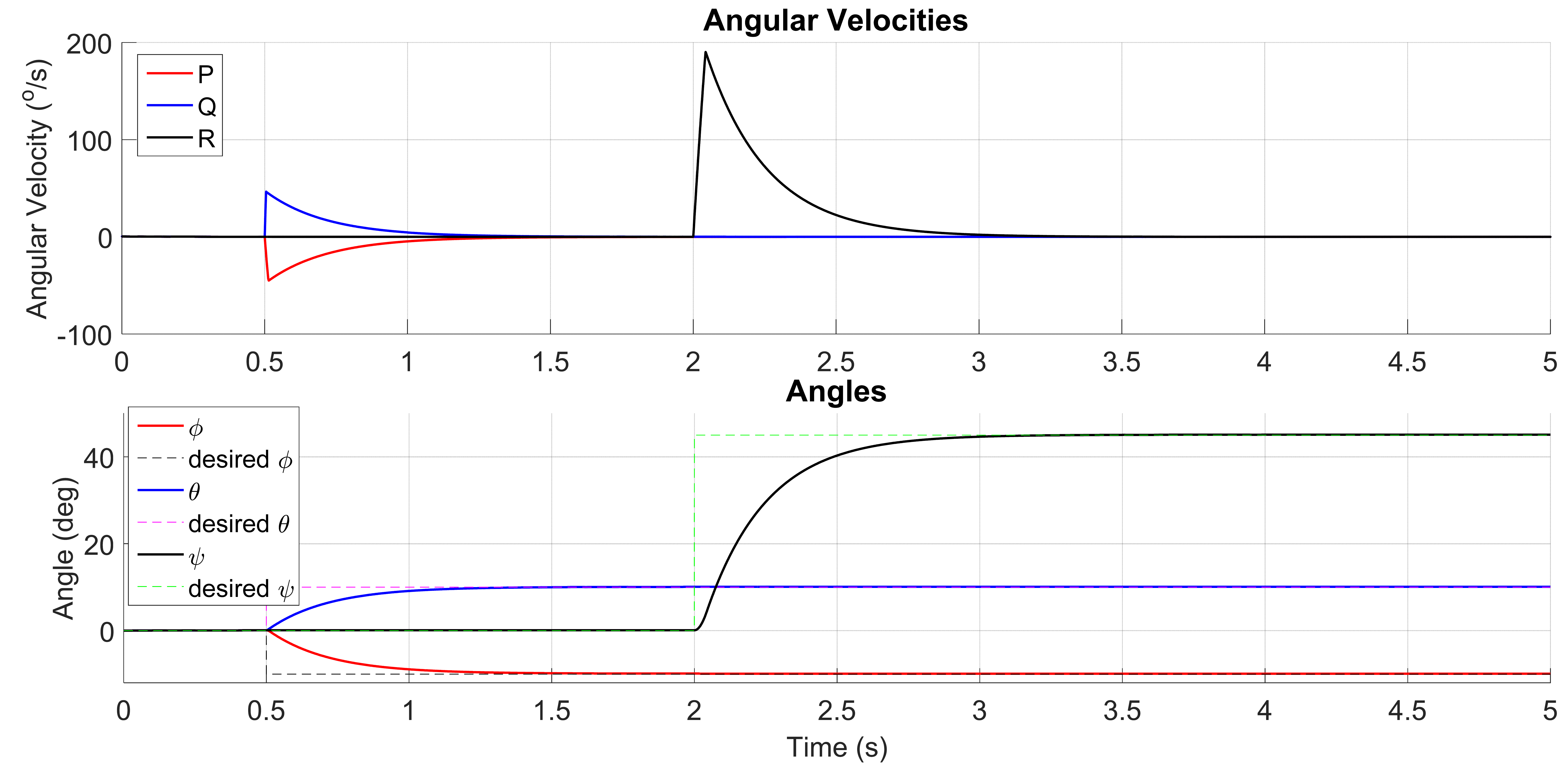}
\caption{Angular velocity and angle responses in the presence of disturbances.}
\label{Fig5}
\end{figure}

In this simulation, a torque disturbance with the amplitude of $0.5 N.m$ is added to all three axes of the quadcopter. The reference values are chosen to be the same as in the previous simulation. The responses are shown in Fig. \ref{Fig5}. As can be seen from the plots, the AQCSM controller can cope with  disturbances to reach the references and maintain the drone stability.

\subsection{Responses to parametric variations}
To evaluate the performance of the proposed controller in different conditions of loads and inertial moments, simulation parameters are varied to tolerate some modelling errors. Specifically, a load of 0.8 kg, corresponding to the maximum payload of the 3DR Solo drone, is added to the model and the following uncertainties are added to the inertial matrix:
\begin{align}
\Delta I = \left[ \begin{array}{ccc}
0 & 0.0044 & -0.0077\\
0.0044 & 0 & 0.0115\\
-0.0077 & 0.0115 & 0\\
\end{array} \right].
\end{align}
\begin{figure}[!htbp]
\centering
\includegraphics[width=8.2cm]{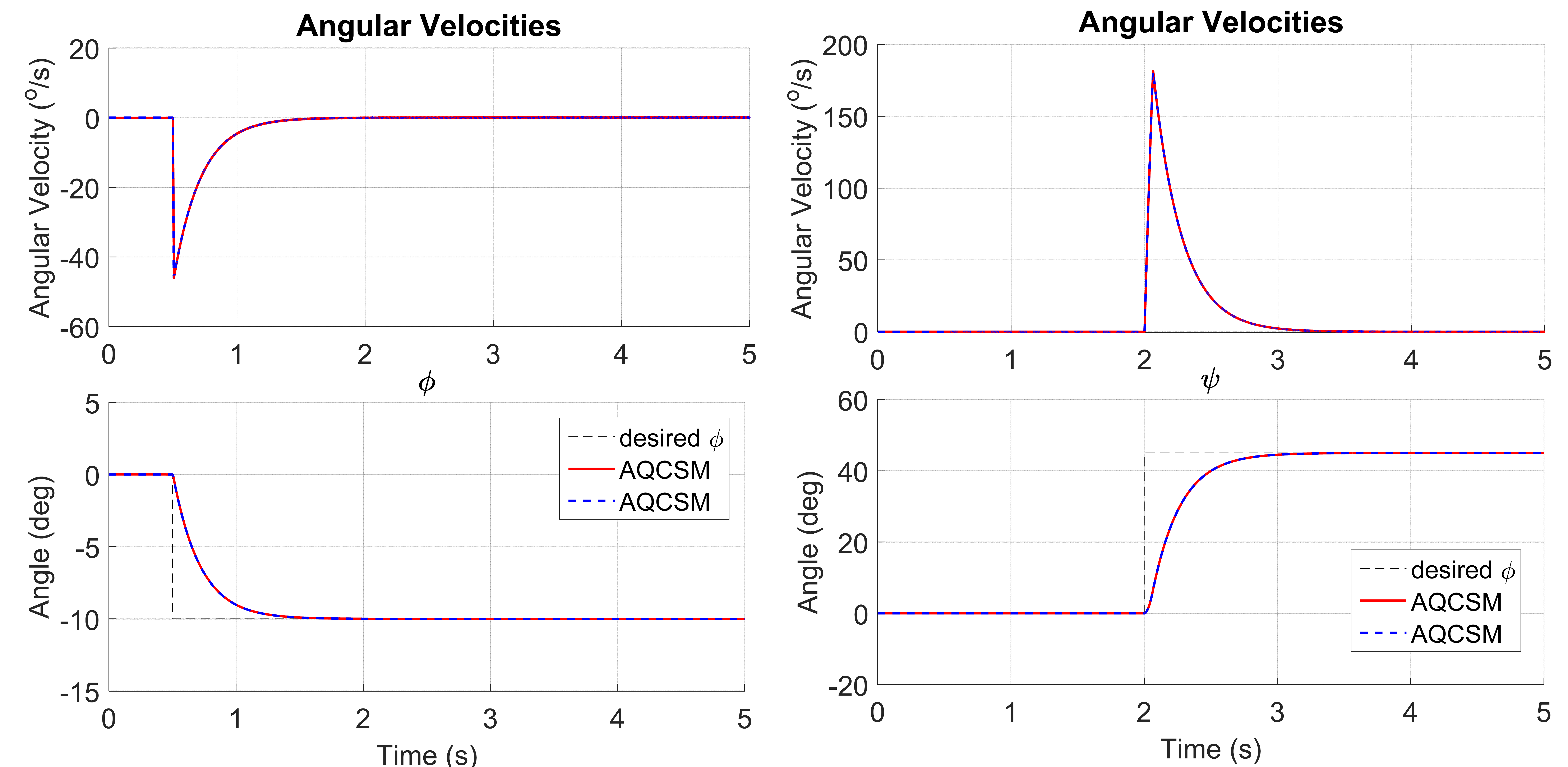}
\caption{Angle and angular velocity responses in the presence of parametric variations.}
\label{vary}
\end{figure}
Figure \ref{vary} shows the results in comparison with the nominal conditions. The almost identical settling time and overshoot between responses corresponding to those scenarios indicates robustness of the proposed AQCSM controller. The adaptive gain $\alpha_1(t)$ response versus time is shown in Fig. \ref{FigAlpha}. The higher gain magnitudes are observed in the two bottom sub-figures imply more energy is required to stabilise the system in dealing with disturbances and uncertainties. This also suggests feasibility of the control scheme.

\begin{figure}[!htbp]
\centering
\includegraphics[width=8.2cm]{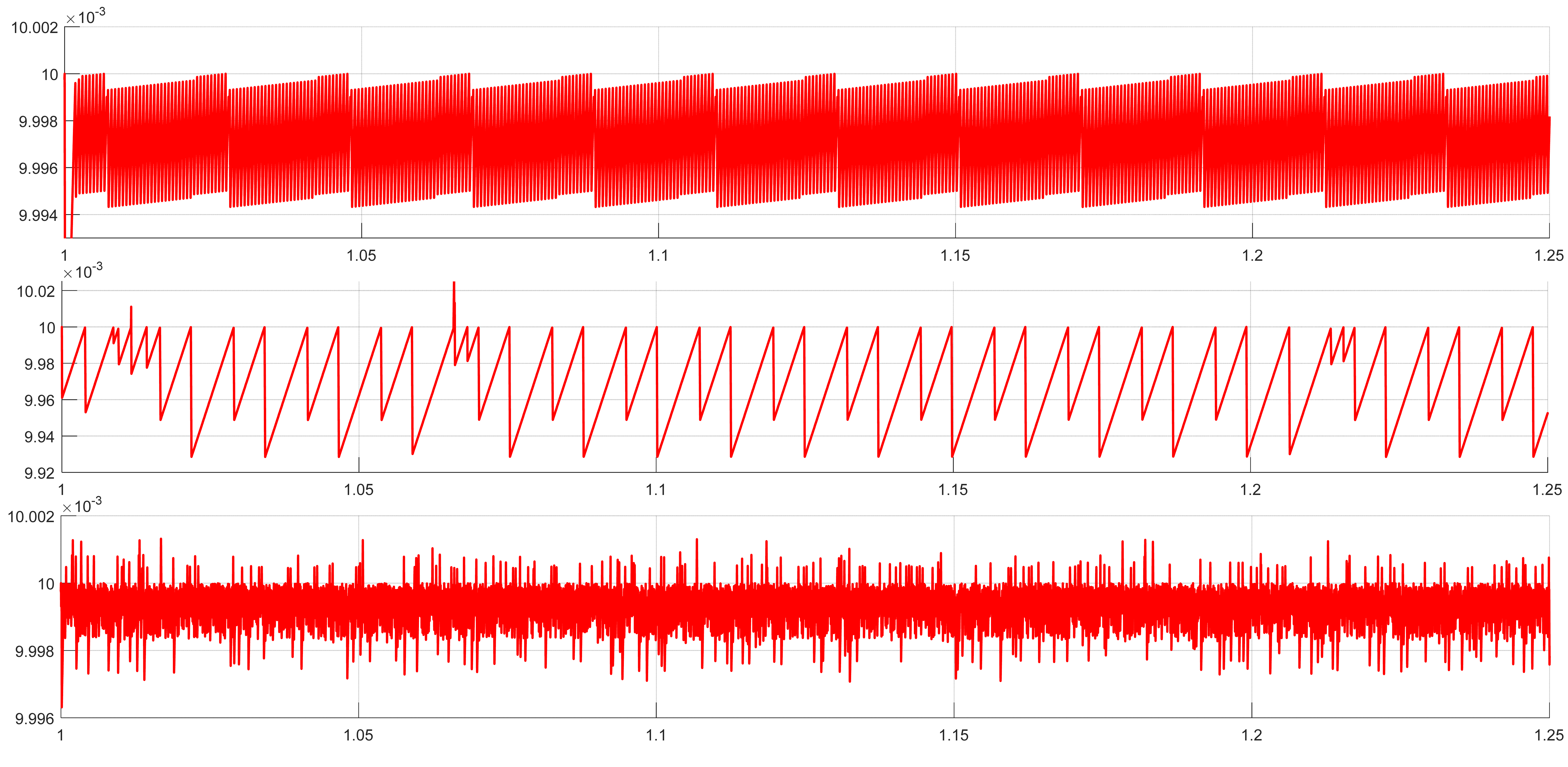}
\caption{The adaptation of gain $\alpha_1(t)$ in various scenarios.}
\label{FigAlpha}
\end{figure}

\subsection{Comparison with real-time data}
\begin{figure}[!htbp]
\centering
\includegraphics[width=8.2cm]{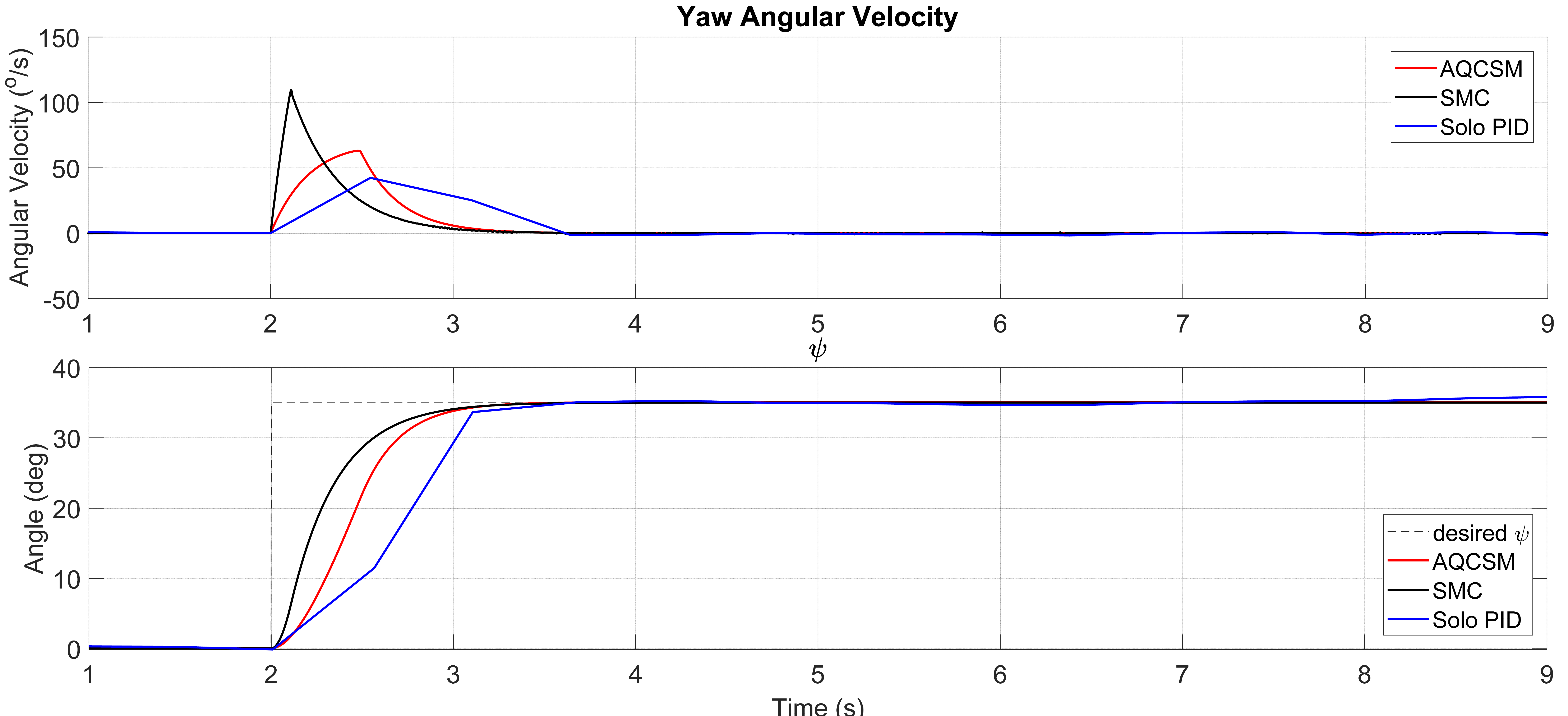}
\caption{Tracking errors - Yaw angular velocity and Yaw angle}
\label{Fig:Comparison}
\end{figure}

To further evaluate the performance of the proposed controller, simulation results are compared with SMC and real time data obtained by using the built-in PID controller of the 3DR Solo drone when performing attitude control during a monitoring task \cite{Ha_2017}. The comparison is carried out by setting the same reference yaw angle to the simulated and real quadcopters. Figure \ref{Fig:Comparison} shows the responses of simulation for AQCSM and SMC as well as experiment for the Solo drone's PID. All controllers reach the reference value without causing much overshoot or oscillation but the AQCSM controller produces better performance with a smoother response. 

\section{Conclusion}
In this paper, an adaptive quasi-continuous sliding mode controller has been developed for robust control of the quadcopters. The control design is based on the selection of a sliding surface and some parameters for adaptation of the control gain taking account into chattering reduction. Control performance is   evaluated in simulation for the cases of both external disturbances and system uncertainties. This robustness property is quite important for civil engineering applications which require accurate attitudes during collecting data for monitoring and inspection tasks. The validity of the proposed control scheme is also judged through comparison with experimental real-time data. Our future work will focus on implementing the proposed controller to develop further high-level planning strategies to take full advantage of UAV-based monitoring and inspection of built infrastructure.

\bibliographystyle{plain}
\bibliography{ISARC}

\end{document}